\documentstyle[emulateapj,psfig,apjfonts]{article}
\def\gs{\mathrel{\raise0.35ex\hbox{$\scriptstyle
>$}\kern-0.6em \lower0.40ex\hbox{{$\scriptstyle
\sim$}}}} 
\def\ls{\mathrel{\raise0.35ex\hbox{$\scriptstyle
<$}\kern-0.6em \lower0.40ex\hbox{{$\scriptstyle \sim$}}}} 
\submitted{Received 2000 -- --; accepted: 2000 -- --}

\lefthead{Smith et al.} 
\righthead{{\it HST} Observations of A\,383}

\begin{document} 
\title{An {\it HST} Lensing Survey of X-ray Luminous Galaxy Clusters:
I. A\,383\footnotemark}

\author{ 
Graham P.\ Smith,$\!$\altaffilmark{2,5}
Jean-Paul Kneib,$\!$\altaffilmark{3}
Harald Ebeling,$\!$\altaffilmark{4,6} 
Oliver Czoske $\!$\altaffilmark{3,5} \&
Ian Smail\altaffilmark{2}
}
\affil{\small 2) Department of Physics, University of Durham, South Road,
Durham DH1 3LE, UK}
\affil{\small 3) Observatoire Midi-Pyr\'en\'ees, CNRS-UMR5572, 14 Avenue
E.\,Belin, 31400 Toulouse, France}
\affil{\small 4) Institute for Astronomy, University of Hawaii, 2680
Woodlawn Drive, Honolulu, HI\,96822, USA}

\setcounter{footnote}{1}

\footnotetext{Based on observations with the NASA/ESA {\it Hubble Space
Telescope} obtained at the Space Telescope Science Institute, which is
operated by the Association of Universities for Research in Astronomy
Inc., under NASA contract NAS 5-26555.} 

\altaffiltext{5}{Visiting Astronomer, Canada-France-Hawaii Telescope
operated by the National Research Council of Canada, the Centre
National de la Recherche Scientifique de France and the University of
Hawaii.}
\altaffiltext{6}{Visiting Astronomer at the W.M.\,Keck Observatory, 
jointly operated by the California Institute of Technology and the
University of California}

\setcounter{footnote}{7}

\begin{abstract} 
We present an analysis of the mass distribution in the core of A\,383
($z=0.188$), one of twelve X-ray luminous galaxy clusters at $z\sim0.2$
selected for a comprehensive and unbiased study of the mass distribution
in massive galaxy clusters.  Deep optical imaging performed by the
{\it Hubble Space Telescope} ({\it HST}\,) reveals a wide variety of
gravitationally lensed features in the core of A\,383, including a
giant arc formed from the strongly-lensed images of two background
galaxies, two radial arcs in the halo of the central cluster galaxy,
several multiply-imaged arcs and numerous arclets.  Based upon the
constraints from the various lensed features, as well as from color
information from ground-based observations, we construct a detailed model
of the mass distribution in the central regions of the cluster, taking
into account both the cluster-scale potential and perturbations from
individual cluster galaxies.  Keck spectroscopy of one component of the
giant arc identifies it as an image of a star-forming galaxy at $z=1.01$
and provides an accurate measurement of the mass of the cluster within
the projected radius of the giant arc (65\,kpc) of $(3.5\pm 0.1)\times
10^{13}$\,M$_{\odot}$.  Using the weak shear measured from our {\it HST}
observations we extend our mass model to larger scales and determine
a mass of  $(1.8\pm 0.2)\times 10^{14}$\,M$_{\odot}$ within a radius
of 250\,kpc.  On smaller scales we employ the radial arcs as probes of
the shape of the mass distribution in the cluster core ($r\ls 20$kpc),
find that the mass profile is more peaked than a single Navarro, Frenk \&
White (NFW, 1997) profile.  Our findings therefore support the proposal
that massive cluster cores contain more mass than can be explained by a
single cluster-scale NFW profile. The optical and X-ray properties of
A\,383 indicate the presence of a central cooling flow, for which we
derive a mass deposition rate of $\gs 200$\,M$_{\odot}$\,yr$^{-1}$.
We also use the X-ray emission from A\,383 to obtain  independent
estimates of the total mass within projected radii of 65 and 250\,kpc:
$(4.0^{+1.1}_{-1.7})\times 10^{13}$\,M$_{\odot}$ and $(1.2\pm 0.5)\times
10^{14}$\,M$_{\odot}$, which are consistent with the lensing measurements.
\end{abstract}



%
%
%
\section{Introduction}

Massive clusters of galaxies represent one extreme of the mass
spectrum of collapsed structures at the present day.  Their properties
are expected to reflect predominantly gravitational processes and 
therefore provide unique insights into the nature and distribution of
the dark matter which drives the formation of structure (e.g.\ Eke et
al.\,1996; Viana \& Liddle 1996; Bahcall, Fan \& Cen 1997; Kay \&
Bower 1999).  In particular, accurate measurements of the mass
distribution in clusters across a range of scales can be used to test
the claim of a universal form for the profiles of dark matter halos
(Navarro, Frenk \& White 1997, NFW) and hence the nature of dark
matter.  At large scales the NFW profile falls off as $\rho \propto
r^{-3}$, steeper than an isothermal model ($\rho\propto r^{-2}$).  On
smaller scales the NFW profile breaks to a shallower slope with
$\rho\propto r^{-1}$, whilst retaining a central cusp.  Analyses of the
mass profiles of dwarf galaxies from dynamical studies have used the
apparent lack of a central cusp in the density distribution to reject
the form of the NFW profile and argue instead for the existence of
self-interacting dark matter (e.g.\ Moore et al.\ 1998).  However, on
larger scales, (e.g.\ luminous elliptical galaxies and massive clusters 
of galaxies) the form of the mass profile has yet to be investigated 
systematically.

The central regions of massive, compact clusters at moderate redshifts
($z\ls 0.5$) act as strong gravitational lenses forming multiple
images of serendipitously placed background galaxies (e.g.\ Smail et
al.\ 1996).  The properties of the images of these background
galaxies (position, relative surface brightness and parity) can be 
used to accurately model the distribution of total mass (both baryonic 
and non-baryonic) within the cluster core (Kneib et al.\ 1996, K96; 
Smail et al.\ 1996; Natarajan et al.\ 1998).  Such mass maps provide 
the most direct and detailed view of the distribution and morphology 
of dark matter in the centers of galaxy clusters.  Strong gravitational 
lensing by rich clusters is therefore an ideal observational tool with 
which to test the halo properties predicted by cosmological models.

The first mass maps constructed from lensing observations were
produced in the early and mid-1990's (Tyson et al.\ 1990; Mellier et al.\ 
1993; Fahlman et al.\ 1994; Kneib et al.\ 1994; Smail et al.\ 1995).  
A significant improvement arrived with the refurbishment 
of {\it HST}\/ which provided the resolution necessary to identify 
faint, lensed features in the crowded cores of rich clusters (e.g.\ K96), 
thereby allowing more detailed mass models to be constructed.

On small scales, unique constraints can be obtained from
radial arcs (rare radially magnified images of background
galaxies) which are found in the very centers of a few cluster lenses.
Such features are very difficult to detect in ground-based
observations, with only a single example known (MS\,2137.3$-$23;
Fort et al.\ 1992; Mellier et al.\ 1993).  In contrast to this, {\it
HST}\/ has uncovered several radial arcs in previously well-studied
cluster lenses (e.g.\ A\,370, Smail et al.\ 1996; AC\,114, Natarajan 
et al.\ 1998) which enabled the first tests of the form of cluster mass
profiles on scales of less than $\sim 100$\,kpc (Williams et
al.\ 1999).

Previous {\it HST}\/ studies of rich clusters covered a heterogeneous
mix of clusters, either selected because they were previously
well-studied (e.g.\ Smail et al.\ 1997) or because they were known to
be strong lenses (e.g.\ K96; Smail et al.\ 1996).  To study the form 
of the mass profile in rich clusters in an unbiased fashion we need 
{\it HST}\/observations of an objectively selected cluster sample.  
Ideally such a sample would be mass-selected; however, in the absence 
of samples compiled from large-scale weak lensing surveys, X-ray 
selected cluster samples (Gioia et al.\ 1990; Ebeling et al.\ 1998, 
2000; DeGrandi et al.\ 1999; Ebeling, Edge \& Henry 2000) are best 
suited to selecting well-defined samples of massive clusters.

Following this premise, we are conducting a survey of twelve of the
most X-ray luminous clusters ($L_{\rm X}\ge 8 \times 10^{44}$\,erg\,s$^{-1}$, 0.1--2.4\,keV) in a narrow redshift slice at $z\sim 0.2$,
selected from the XBACs sample (X-ray Brightest Abell-type Clusters;
Ebeling et al.\ 1996).  As XBACS is restricted to Abell clusters (Abell, 
Corwin \& Olowin 1989), it is X-ray flux limited and not truly X-ray 
selected.  However, a comparison with the X-ray selected {\it ROSAT}
\/ Brightest Cluster Sample (BCS, Ebeling et al.\ 1998, 2000) shows 
that $\sim 75$\% of the BCS clusters in the redshift and X-ray 
luminosity range of our sample are in fact Abell clusters.  Hence, 
our XBACs sample is, in all practical aspects, indistinguishable 
from an X-ray selected sample.

In this paper we describe {\it HST}\/ observations of the core of one
of the first clusters observed in our survey, A\,383 ($z=0.188$),
which reveal a multitude of strongly-lensed features.  In combination
with color information obtained from ground-based imaging
observations in three passbands, the lensed features are used
to produce an accurate model of the mass distribution within the
central 500\,kpc of the cluster. We complement these results with
an analysis of the X-ray properties of A\,383 using archival {\it ROSAT} 
{\it HRI} data. In \S2 we describe the observational data, their reduction 
and analysis, followed in \S3 by a discussion of the interpretation 
of these data and the construction of the lens model of the cluster.  
Finally in \S4 we summarize the main results of our analysis and 
present our conclusions.  We adopt $H_0=50$\,kms$^{-1}$\,Mpc$^{-1}$ and 
$q_0=0.5$ throughout.  At the cluster redshift $1''\equiv 4.0$\,kpc in 
this cosmology.

\section{Data Reduction and Analysis}

A\,383 is a massive cluster of galaxies at $z=0.188$ with a core
dominated by a luminous cD galaxy (Figs.~1 \& 3).  The X-ray luminosity
from a pointed observation with the {\it ROSAT} {\it HRI} is $(9.8\pm
0.3)\times 10^{44}$\,erg\,s$^{-1}$ in the 0.1--2.4\,keV band (see 
\S2.4).  

\subsection{{\it HST} Imaging}

A\,383 was observed with the {\it HST} {\it WFPC2} camera on January 25,
2000.  Three exposures totaling 7.5\,ks were taken through the F702W
filter.  Each exposure was shifted relative to the others by 10 WFC
pixels ($1.0''$) providing a partial overlap of the chip fields.
After pipeline processing, standard {\sc iraf/stsdas} routines were
employed to shift and combine the frames to remove both cosmic rays
and hot pixels.  Corrections for under-sampling of the point spread
function and geometric distortion of the optics were made using the
{\sc dither} package within {\sc iraf} (Fruchter \& Hook 1997).  The
final frame (Fig.~1) has a pixel scale of 0.05$''$, an effective
resolution of 0.15$''$ and a 1$\sigma$ detection limit within 
the seeing disk of $R_{\rm 702}\simeq 31$.

To produce a catalog of faint arclets from our data we first analysed
the {\it HST} frame using the SExtractor package (Bertin \& Arnouts
1996).  All objects with isophotal areas in excess of 30 pixels
($0.074$\,arcsec$^{2}$) at the $\mu_{702} = 24.5$\,mag\,arcsec$^{-2}$
isophote ($3\sigma$/pixel) were selected.  A steep roll-over in the
observed differential number counts arising from incompleteness occurs
at $R_{\rm 702}\sim 25.5$.  We estimate the completeness to be $\sim80$\%
(approximately 5$\sigma$) at $R_{\rm 702}\sim 25.5$, which we adopt as our
magnitude limit, giving a total of 457 sources within a 4.9 arcmin$^{2}$
area (excluding the PC chip).  Seventeen of these sources are classified
as star-like on the basis of their profile shapes and are excluded from
our analysis.

For the purposes of the photometric analysis presented in Table~1 we
correct the F702W photometry ($R_{\rm 702}$) to the Cousins $R$-band
in the following manner.  The observed range in $(B-R)$ colors of the
arcs roughly translates to $0.2<(V-R)<1.0$ giving a typical color of
$(V-R)\sim 0.6$. Adopting this color we obtain a correction of $R_{\rm
702}-R=0.2$ (Holtzman et al.\ 1995), and estimate that this correction
introduces a systematic uncertainty in the $R$-band photometry of $\pm
0.06$ magnitudes.

The final {\it HST} frame (Fig.~1) reveals many previously unknown
strongly lensed features, including a giant arc and two radial arcs, 
making A\,383 a striking new addition to the catalog of cluster lenses 
at intermediate redshifts.

%
%
\begin{figure*} 
\vspace*{3.0in}
\centerline{\Huge fig1.gif}
\vspace*{3.0in}
\vspace{0.2cm}
\noindent{ \addtolength{\baselineskip}{-3pt} {\sc Fig.\,1.} -- The
central region of A\,383 as observed with {\it HST} WFPC2, overlaid
with an isodensity representation of our lens model.  Contours
correspond to projected surface mass densities of 3, 4, 6, 8, 11, 15$
\times 10^{9}$\,M$_{\odot}$\,kpc$^{-2}$.  The numerical labels indicate
the cluster members used in the lens model.  The alpha-numeric labels
identify the multiple images used to constrain the lens model and a
number of other singly-imaged arclets.  The spiral galaxy labeled B16
was included in our spectroscopic sample and has $z=0.6560$.  The 
disturbed morphology of the central galaxy, including what appears to 
be a dust lane passing between two peaks in the light profile, is 
apparent in the center of this figure.

\addtolength{\baselineskip}{3pt} }

\vspace{0.2cm}

\end{figure*}

\subsection{Ground-based Imaging}

With the aim of extending our lensing mass map to the turn-around radius
of the cluster, we have used the 3.6m Canada-France-Hawaii Telescope
(CFHT) with the CFH12K camera to obtain panoramic images of A\,383 on
the nights of November 14--16, 1999.  Total exposure times of 7.2\,ks,
6.0\,ks and 3.6\,ks, accumulated at 6--10 dither positions, were
acquired in the $B$, $R$ and $I$ bands respectively.  Data reduction
was performed within {\sc iraf} using the {\sc mscred} package
including standard bias subtraction and flat-fielding using twilight
flats.  The dithered exposures were aligned with the Digital Sky
Survey frame of the same field to an rms accuracy of 0.15$''$.  More
information on the reduction and analysis of these observations will
be presented in a forthcoming paper (Czoske et al.\ 2000).  Here we
use the central regions of the $B$ and $I$ band frames to provide
photometry of the lensed features in the cluster core (see Table~1).
These two frames have seeing of 0.88$''$ and 0.71$''$ FWHM
respectively.

In addition to the CFH12K imaging, we have obtained $K$-band images 
of the core of A\,383 with the UFTI imager on the 3.8m United Kingdom 
Infrared Telescope (UKIRT), Mauna Kea, on October 14, 1999.  The final 
frame was accumulated in 27 dithered sub-exposures of 90\,s duration 
each to give a total on-source integration time of 2.4\,ks, all in 
photometric conditions. Employing standard procedures these frames were 
reduced, combined and calibrated using observations of UKIRT faint 
standards bracketing the science exposures.  The final frame has seeing 
of 0.42$''$ FWHM with 0.09$''$/pixel sampling and an effective 5$\sigma$ 
depth of $K=20.3$.

Optical and optical-infrared colors of the lensed features in 
A\,383 were measured off the aligned and seeing-matched $BRIK$ frames 
using apertures designed for individual arcs.  The arcs were first 
masked out of the science frame and the sky background 
estimated by median smoothing over the aperture defined by the absent 
arc.  The sky subtraction was then performed and the individual arc 
apertures applied to the sky-subtracted frame to obtain the $BRIK$ 
photometry presented in Table~1.  Error bars were estimated by varying 
the smoothing length used in estimating the sky background.

\subsection{Spectroscopy}

On January 26, 2000 we observed the cD galaxy in A\,383 with
the Wide Field Grism Spectrograph on the University of Hawaii's 2.2-m
telescope.  The low dispersion spectrum obtained in a 1.8\,ks exposure
in extremely poor seeing (2.7$''$) shows the full range of emission
lines typically found in cD galaxies in cooling flow clusters (e.g.\
Crawford et al.\ 1999) and places the cD at a heliocentric redshift of
$z=0.1880\pm 0.0012$.

On January 29, 2000 we obtained 28 additional spectra with the LRIS
spectrograph (Oke et al.\ 1995) on the Keck-II 10-m telescope in MOS
mode, with a total exposure time of 3.6\,ks in average seeing of
0.9$''$.  Use of the 300/5000 grating centered at 7500\,\AA\ provided
wide spectral coverage (5000--10,000\,\AA) at a spectral resolution of
2.55\,\AA/pixel.  Redshifts were measured independently by three members 
of our team (OC, HE, JPK) and also using the {\sc rvsao} cross-correlation 
package under {\sc iraf} to accurately estimate the errors.  Successive 
$3\sigma$ clipping around the peak of the radial velocity distribution 
yields a redshift of $z=0.1880\pm0.0002$ for A\,383 and a velocity 
dispersion of $\sigma = 1150^{+150}_{-200}$ kms$^{-1}$ from 18 galaxy 
redshifts.  Ten other galaxies were found to be background, including 
B16 at $z=0.656$, which lies close to the cluster core (see Fig.~1).

%
%
\vspace*{3mm}
\centerline{\psfig{file=fig2.ps,width=3.5in,angle=270}}
\noindent{\addtolength{\baselineskip}{-3pt} \scriptsize {\sc Fig.\,2.}
-- The spectrum of arc B0a in A\,383 taken with the LRIS spectrograph
on the Keck-II 10m telescope.  The arc exhibits a blue continuum and
we identify the strong emission line as [O{\sc ii}]$\lambda$3727 and
confirm this with several UV absorption features (marked) to give a
redshift of $z=1.0103\pm 0.0001$.  The upper spectrum is smoothed to
the nominal resolution of the spectrograph and offset vertically for
clarity.  The shaded regions show areas of the spectrum strongly
affected by night sky lines.

\addtolength{\baselineskip}{3pt} }
\vspace*{3mm}

We also obtained a spectrum of the lensed feature B0a which we show
in Fig.~2.  We identify the strong emission line at 7493\AA\ as
[O{\sc ii}]\,$\lambda$3727 which places the galaxy at a redshift
of $z=1.0103\pm0.0001$.  This interpretation is confirmed by the
identification of Fe{\sc ii}\,$\lambda$2600 and Mg{\sc ii}\,$\lambda$2800
absorption features.  The properties of this galaxy are discussed further
in \S3.

\subsection{{\it ROSAT/HRI} Observation}

A\,383 was observed with the {\it ROSAT} {\it HRI} in February 1995 for a
total integration time of 27.4\,ks (ROR \#800890).  An adaptively 
smoothed map of the observed X-ray emission (created using the {\sc 
asmooth} algorithm of Ebeling, White \& Rangarajan 2000) is shown in 
Fig.~3.  We used a modified version of Steve Snowden's {\sc cast hri} 
software to correct for exposure time variations across the 
field of view and subtract a particle background component.  

~From the fully processed {\it ROSAT} {\it HRI} image the total
background-corrected {\it HRI} count rate from A\,383 is  measured
to be $(0.128\pm 0.005)$\,ct\,s$^{-1}$ within 1.5\,Mpc (6.2$'$) of the
cluster center, using the mean diffuse background at $r>2$\,Mpc (8.2$'$).
Assuming a standard plasma emission spectrum,  a Galactic hydrogen column
density of $4.1\times 10^{20}$\,cm$^{-2}$ for this sight-line (Dickey \&
Lockman 1990), a metallicity of 0.3 and an ambient X-ray temperature of
$\sim 7.1$\,keV (estimated from the cooling-flow-corrected cluster $L_{\rm
X}$--$kT$ relation of Allen \& Fabian 1998) we derive a total unabsorbed
cluster flux of $(6.55\pm 0.23)\times 10^{-12}$\,erg\,cm$^{-2}$\,s$^{-1}$
(0.1--2.4\,keV).  The corresponding X-ray luminosity is $(9.8\pm
0.3)\times 10^{44}$ erg s$^{-1}$ (0.1--2.4 keV),  in agreement with the
XBACs value of $(8.0 \pm 2.4)\times 10^{44}$\,erg\,s$^{-1}$ in the same
energy band (Ebeling et al.\ 1996).

%
%

\vspace*{0.8in}
\centerline{\Huge fig3.gif}
\vspace*{0.8in}

~\bigskip
\noindent{\scriptsize \addtolength{\baselineskip}{-3pt} 

{\sc Fig.\,3.}
-- X-ray flux contours of the adaptively smoothed emission from A\,383
as seen with the {\it ROSAT} {\it HRI}, overlaid on an $R$-band image of
the cluster taken with the CFH12K camera (see \S2.2).  All X-ray features
are at least $3\sigma$ significant with respect to the local background.
The lowest contour lies 30\% above the background, adjacent contours
differ by 30\%. We adjusted the astrometry of the HRI image by 4$''$
(well within the errors of the astrometry solution) to align the X-ray
peak with the cD galaxy. The near circular symmetry of the X-ray flux
contours reflects the symmetry of the lensing mass contours plotted
in Fig.~1.  The region shown in our Fig.~1 is marked by the dashed box.

\addtolength{\baselineskip}{3pt} }

\section{Modelling and Results}

In this section we first describe and interpret the multiply-imaged
features observed in our {\it HST} frame of A\,383 and explain how
these  were used to constrain the mass distribution in the central
regions of the cluster.  We then discuss the form of the mass profile
we derive in the very center of the cluster and compare this to
theoretical expectations from high resolution N-body simulations.
Finally we discuss the X-ray properties of A\,383 and compare the mass 
distribution determined from the lens model with the estimates based 
on the {\it ROSAT} data.

\subsection{Multiply Imaged Features}

As a result of the superb resolution of our {\it HST} images, a detailed
analysis of the morphology of the observed arcs allows us to identify
seven multiply-imaged systems in the field of A\,383.  This is 
significantly greater than the number typically seen in ground-based 
observations of clusters.  We discuss each multiply-imaged system below, 
and summarize their photometric and spectroscopic properties in 
Table~1.  The lensed features are identified in Fig.~1 and 
the multiple images are further illustrated below in Fig.~4--6.

\subsubsection{The Giant and Radial Arcs}

We interpret the giant and radial arcs in A\,383 as being the lensed
images of two background galaxies, B0 and B1, at $z=1.01$ (\S2.3) and
$z\sim1.1$ respectively.  This interpretation arises from modelling
(\S3.2) of the radial arcs, which constrains the outer radial arc (B1d)
to be a counter image of B1a/b/c at $z\sim1.1$, and the inner radial arc
(B0b) to be a counter image of B0a.  B0a is also slightly closer to the
center of the cluster than other components of the giant arc, supporting
the suggestion that it is at a slightly lower redshift than B1.

\vspace*{0.3cm}
%
%
\centerline{\psfig{file=fig4.ps,width=3.5in,angle=-90}}

\noindent{\scriptsize \addtolength{\baselineskip}{-3pt} {\sc Fig.\,4.}
-- The giant arc in A\,383 (see text for interpretation).  Each tick
mark represents $1''$, and the orientation is as in Fig.~1.  The dotted lines indicate the path of the $z=1.1$ critical line close to the observed arcs.

\addtolength{\baselineskip}{3pt} }

\vspace*{0.2cm}

{\it B0 (Figs.~4 \& 5)} --- This system comprises the brightest component
of the giant arc (B0a) and the inner of the two radial arcs (B0b).
It is also plausible that B15 is a component of the B0 system given
its proximity to B0a in the image plane, however it's redder color
(Table~1) suggests that it is either singly imaged, or that B0b has a
significant color gradient.  We are unable to measure the color of B0b
to confirm this possibility due to the light from the central galaxy.
We have also estimated the unlensed $R$-band magnitude of B0 (Table 1)
which, along with it's blue optical-infrared color, suggests that B0 is
a low luminosity star-forming galaxy ($M_V\sim -19.3$).

\vspace{0.2cm}

%
%
\centerline{\psfig{file=fig5.ps,width=2.0in,angle=-90}}

\noindent{\scriptsize \addtolength{\baselineskip}{-3pt} {\sc Fig.\,5.}
-- A view of the radial arcs as seen on the {\it HST} frame (enhanced
by subtracting a median-smoothed frame from the science frame).  Each
tick mark represents $1''$, and the orientation is as in Fig.~1.

\addtolength{\baselineskip}{3pt} }

\vspace{0.2cm}

{\it B1 (Figs.~4 \& 5)} --- B1a/b is a pair of merging images straddling
the $z\sim1.1$ critical line with counter-images B1c, at the extreme right
hand end of the giant arc and B1d, the outer of the two radial arcs.
The lens model (\S3.2) and photometric analysis (Table~1) support
this interpretation and appear to exclude B15 from being a further
counter-image.  The unlensed $R$-band magnitude of B1 has been estimated
(Table~1), from which we determine that it is a very low luminosity galaxy
($M_V\sim -18.3$).

\subsubsection{Perturbations by Galaxy Halos}

{\it B2 and B3 (Fig.\,6)} --- The lensing effect of neighboring
cluster galaxy halos on this complex group of blue arcs demonstrates
how galaxy-scale masses measurably perturb the cluster potential.  We
have modeled the effect of these perturbations and resolved B2 and B3 
into two background galaxies, both at a redshift of $z\sim 3.5$.  
Although the photometric analysis is hampered by the nearby galaxy halos, 
the data presented in Table~1 lend additional support to this
interpretation.

B2 comprises five images, of which B2a/b and B2d/e are pairs of images
straddling the $z\sim 3.5$ critical line as it passes around the halos
of cluster ellipticals \#46 and \#23 respectively.  B2c lies outside
this critical line and has the same parity as B2a.  A sixth candidate
counter image lies approximately 1$''$ to the left of B3b under the
halo of cluster elliptical \#23.  Multi-band {\it HST} imaging is 
necessary to achieve the accurate photometry required to confirm this 
hypothesis.  There is also a low signal-to-noise feature lying in the 
saddle region between cluster ellipticals \#46 and \#76, which is 
probably an extension of B2c.  Counter images of this portion of the 
B2 system are undetected, due to the intrinsic faintness of the galaxy 
and the relatively low magnification of the counter images relative to 
the influence of the saddle on the observed feature.  Finally, we 
estimate that the B2 background galaxy has $M_V\sim -22.0$ and the blue 
colors typical of a star-forming galaxy.

\vspace{0.2cm}

%
%
\centerline{\psfig{file=fig6.ps,width=3.5in,angle=-90}}

\noindent{\scriptsize \addtolength{\baselineskip}{-3pt} {\sc Fig.\,6.}
-- The B2/B3 system of arcs (see text for more details).  Cluster
ellipticals \#46, \#76 and \#23 are also labelled.  Each tick mark
represents $1''$, and the orientation is as in Fig.~1.  The dotted 
lines indicate the path of the $z=3.5$ critical line close to the 
observed arcs.

\addtolength{\baselineskip}{3pt} }

\vspace{0.2cm}

B3 consists of three images; B3b/c is adjacent to cluster elliptical 
\#23, and B3a, the counter image, lies outside the critical line, 
$\sim 1''$ below and to the right of B2c in Fig.~6.  B3d appears to be a 
singly-imaged component of the background galaxy that produces B3a/b/c.  
This interpretation is supported by the photometric analysis in Table~1.  
We also estimate that B3 is a star forming galaxy with a similar 
luminosity ($M_V\sim -22.4$) to B2.

{\it B4 (Fig.\,1)} --- This is another complex and unusual arc,
the precise nature of which is uncertain.  One possibility is that B4 
is part of a multiple image system containing B6; however, photometric 
analysis (Table~1) does not support this hypothesis.  More 
complete color information will improve our ability to comment on 
this arc.  If B4 is a singly imaged source it lies at a redshift 
of $z\ls 1.0$.

\subsubsection{Faint Arcs}

{\it B5 (Fig.\,1)} --- B5 is a very faint arc lying in the saddle
region between the central galaxy and cluster elliptical \#99.  We
interpret this arc as an image of a background galaxy which is magnified 
by this saddle and therefore constrain the redshift of the galaxy to 
be $z\gs 2.4$.

{\it B9a \& B9b (Fig.\,1)} --- This is a pair of extremely faint arcs
detected only because of the superb resolution and low sky background
of the {\it HST} observations.  Both arcs appear to comprise three
images resulting from the magnifying effect of the saddle potential 
between the cluster center and cluster elliptical \#260.  We estimate 
that they lie at a redshift of $z\gs 3$.

{\it B14 (Fig.\,1)} --- B14 is barely detected above the sky
background in the {\it HST} frame, however we detect it at 
$K=19.70\pm 0.05$ in the UKIRT frame, giving $(R-K)=6.02\pm 0.11$.  
As B14 is singly-imaged, we obtain a redshift limit of $z\ls 3.9$ 
from our lens model.  A full analysis of galaxies exhibiting 
extreme colors and the general background population lying behind 
our cluster sample will be presented in a future paper.

\subsubsection{Arclets}

We also detect numerous singly-imaged arclets in the field of 
A\,383 (Fig.~1 and Table~1) and have obtained upper limits on 
the redshifts of three of these arclets (B6, B7 and B8) on the 
basis of their singly-imaged nature.

{\it B6 (Fig.\ 1)} --- We obtain a redshift limit of $z\ls0.9$ 
for this arclet, however it's blue $(B-R)$ and $(R-I)$ colors (Table~1) are consistent with it being a star forming galaxy at 
$z \sim 1.5$--2.0.  The latter interpretation supports the 
possibility that B6 and B4 contain multiply imaged features 
of a single background galaxy.  As noted in \S3.1.2, improved 
photometry of B4 will help to resolve this uncertainty.

{\it B7 (Fig.\ 1)} --- B7 lies at a redshift of $z\ls 1.3$ which, 
on the basis of its $(B-R)$ color (Table~1) is consistent with B7 being
a moderately star forming galaxy.

{\it B8 (Fig.\ 1)} --- This arclet has a redshift limit of $z\ls 2.0$,
while its colors are broadly consistent with this being a star forming 
galaxy at  $z \sim 1.0$--1.4.

%
%
{\scriptsize

\begin{table*}
\begin{center}
\caption{\centerline {Photometry and Spectroscopy of Lensed Features in A\,383}}
\vspace{0.1cm}
\begin{tabular}{lccccccc}
\hline\hline
\noalign{\smallskip}
{Feature} &  $R$ & $(B-R)$ & $(R-I)$ & $(I-K)$ & $z$ &  $R_{\rm source}^{d}$\cr
          &      &       &       &       &           \cr
\hline
\noalign{\smallskip}
B0a  & $21.98\pm0.01$ & $1.02\pm0.01$ & $0.74\pm0.01$ & $1.91\pm0.03$ & $1.0103\pm0.0001^a$  &  $25.0\pm0.3$\cr
B0b  & $21.95\pm0.24$ &  ... & ... & ... & \cr
\noalign{\smallskip}
B1a/b  & $21.50\pm0.02$ & $1.14\pm0.03$ & $1.07\pm0.04$ & $2.06\pm0.14$ & $1.1\pm0.1$ & $26.7\pm0.6$ \cr
B1c  & $23.03\pm0.02$ & $1.11\pm0.03$ & $1.00\pm0.03$ & $1.90\pm0.07$ & \cr
B1d  & $23.94\pm0.07$ & ... & ... & ... &  \cr
\noalign{\smallskip}
B2a/b  & $23.79\pm0.06$ & ... & ... & ... & $3.5\pm0.2$  & $26.8\pm0.7$ \cr
B2c  & $23.90\pm0.01$  &  $0.87\pm0.03$  &  $0.29\pm0.02$  &  $1.94\pm0.08$ \cr
B2d/e  & $22.56\pm0.21$  &  $0.75\pm0.26$  &  $0.54\pm0.23$  &  $<2.3$  \cr
\noalign{\smallskip}
B3a  & $23.80\pm0.02$  &  $0.83\pm0.02$  &  $0.40\pm0.02$  &  $<1.9$ &  $3.5\pm0.2$ &  $26.3\pm0.2$ \cr
B3b/c  & $22.62\pm0.24$  &  $0.66\pm0.27$  &  $0.59\pm0.27$  & $<2.5$  \cr
B3d  & $23.86\pm0.02$  &  $1.09\pm0.03$  &  $0.50\pm0.02$  &  $<2.0$ & \cr
\noalign{\smallskip}
B4  & $22.07\pm0.01$  &  $0.94\pm0.26$  &  $<0.4$  &  ... & $\ls1.0^{b}$ & ... \cr
\noalign{\smallskip}
B5  & $23.35\pm0.03$  &  ... &  ...  &  ...  & $\gs2.4$ & ...\cr
\noalign{\smallskip}
B6  & $23.42\pm0.01$  &  $0.42\pm0.03$  &  $0.12\pm0.07$  &  $<3.5$  & $\ls0.9^{b}$ & ... \cr
\noalign{\smallskip}
B7  & $24.23\pm0.02$  &  $0.63\pm0.02$  &  $<0.5$  &  ... & $\ls1.3^{b}$ & ... \cr
\noalign{\smallskip}
B8  & $24.68\pm0.05$  &  $0.84\pm0.14$  &  $0.81\pm0.17$  & $<4.1$ & $\ls2.0^{b}$ & ... \cr
\noalign{\smallskip}
B9a  & $25.20\pm0.11$  & ...  &  ... &  ...  & $>3.0$ & ... \cr
B9b  & $24.48\pm0.08$  & ...  &  ... &  ...  & $>3.0$ & ... \cr
\noalign{\smallskip}
B10  & $24.85\pm0.05$  &  $0.19\pm0.03$  &  $0.57\pm0.07$  & $<4.5$ & ... & ... \cr
\noalign{\smallskip}
B11  & $22.92\pm0.01$  &  $1.60\pm0.01$  &  $1.02\pm0.01$  & $2.46\pm0.01$ & ... & ... \cr
\noalign{\smallskip}
B12  & $23.56\pm0.02$  &  ...  &  ...  &  ... & ... & ...   \cr
\noalign{\smallskip}
B13  & $23.97\pm0.01$  &  $1.45\pm0.14$  &  $1.05\pm0.16$  & $2.19\pm0.18$ & ... & ... \cr
\noalign{\smallskip}
B14$^c$  & $26.12\pm0.02$  &  $>1.2$ &  $<2.0$  &  $>4.0$  & $\ls 3.9$ & ... \cr
\noalign{\smallskip}
B15  & $22.23\pm0.02$  &  $1.35\pm0.01$  &  $0.94\pm0.02$  &  $1.96\pm0.09$ &  $1.1\pm0.1$ & ... \cr
     &                 &                 &              &            \cr
\hline
\noalign{\smallskip}
$L_{*}$ Cluster E/S0  & 18.01 & 2.44 & 0.67 & 2.50 & 0.188 \cr
\noalign{\smallskip}
\noalign{\hrule}
\noalign{\smallskip}
\end{tabular}
\end{center}

a) Spectroscopic redshift; b) Redshift assumes single image; c)
Extremely Red Object with $(R-K)=6.02\pm0.11$.  d) Estimated $R$-band
magnitude of the background galaxy, corrected for lens amplification;
the error bars reflect the uncertainties in the lens model and generally
scale with the number of multiple images per background galaxy. 

\vspace*{-0.3cm}
\end{table*}
}

\subsection{Lens Model}

The mass distribution of A\,383 was reconstructed using the parametric
lens inversion method described in detail by K96.  In addition to a
cluster-size mass component, the model contains 29 individual galaxy
masses (including the central galaxy) covering the entire {\it HST}
field of view.  The parameters describing the model were constrained
were constrained within the cluster core using four of the multiple
image systems identified in the {\it HST} frame (B0, B1, B2 and B3)
and on larger scales using the weak shear field estimated from the
shapes of the faint galaxies across the whole frame.  The cluster
scale mass that dominates the mass distribution has a core radius of
($46\pm 3$)\,kpc, velocity dispersion of ($900\pm 20$)\,kms$^{-1}$ and a
cut-off radius of ($1200\pm 100$)\,kpc (see K96 for details of how these
quantities are defined).

The mass distribution obtained from the lens model is overlaid on
the optical image in Fig.~1.  We calculate the total projected mass
within the radius of the giant arc (65\,kpc) to be $(3.5\pm 0.1) \times
10^{13}M_{\odot}$.  Using the weak shear measurements we can extend our
model to larger scales and constrain the projected mass within a 250\,kpc
radius of the cluster center to be: $(1.8\pm0.2) \times 10^{14}M_{\odot}$.

In Fig.~7 we plot the projected surface mass density derived
from the lens model as a function of the distance from the cluster center
($r$).  On scales $r\gs 10$ kpc, the density profile is dominated by
the cluster-scale potential, the core radius of the cluster-scale mass
distribution ($\sim 46$\,kpc) being consistent with previous estimates of
core radii from lensing studies (Mellier et al.\ 1993; Smail et al.\ 1996;
Natarajan et al.\ 1998).  The shape of A\,383's profile is however only
constrained out to $r\sim 400$\,kpc, this being the extent of the {\it
WFPC2} field of view.   The mass profile is most accurately measured
within $r\sim 100$\,kpc where information from the multiple images
is available.

%
%
\vspace*{3mm}
\centerline{\psfig{file=fig7.ps,width=3.0in,angle=-90}}

\noindent{\scriptsize\addtolength{\baselineskip}{-3pt} {\sc Fig.\,7.}
-- The projected surface mass density ($\Sigma$) of A\,383 as a
function of radius, obtained from the lens model by measuring the 
mass density in concentric annuli.  The dashed lines illustrate the 
difference between the profile slope at the positions of the two
radial arcs (B0b and B1d), which are marked by arrows at 6.0 and 19.9\,kpc respectively.

\addtolength{\baselineskip}{3pt} 
}
\vspace*{3mm}

\subsubsection{Radial Arc Properties}

B0b and B1d (see Figs.~1 \& 5) lie at distances of ($6.0\pm 0.4$)\,kpc and
($19.9\pm 0.4$)\,kpc from the cluster center respectively.  These two arcs
therefore enable us to probe the shape of the mass distribution on
scales of $r\ls 20$\,kpc.  Specifically, the radial arcs constrain the
slope of the density profile, as their radial positions depend on the 
local gradient of the projected mass, rather than on the mean 
enclosed surface density.

A detailed analysis of the shape of A\,383's density profile is currently
limited by the pseudo-isothermal elliptical functional form adopted in
the lens model (see K96).  However, in order to estimate the slope of the
density profile at the positions of the two radial arcs, we fitted various
polynomial functions to the density profile data.  De-projecting this
analysis to obtain the slope of the three-dimensional density profile
($\rho$), we obtain the gradients: $d(\log\,\rho)\,/\,d(\log\,r)=-1.41$
and $-1.29$ at $r=6.0$\,kpc (B0b) and $r=19.9$\,kpc (B1d) respectively;
the error on these slopes is of the order $\ls 0.03$.

We interpret this difference in slope at the positions of the two radial
arcs, specifically the steeper slope at $r\sim 6$\,kpc, as being due
to the presence of a distinct central galaxy halo within the overall
cluster halo.

Williams et al.\ (1999) suggest that radial arcs are a useful test
of the lensing role of such central galaxy halos within an overall
cluster-scale NFW halo.  They predict that $\log(\theta_{r}/\theta_{t})$
(where $\theta_{r}$ and $\theta_{t}$ are the angular positions of the
radial and tangential arcs respectively) should decrease with decreasing
cluster mass as the lensing role of the central galaxy becomes more
significant relative to the overall cluster mass distribution.  From their
Fig.~7 we estimate that for A\,383: $\log(\theta_{r}/\theta_{t})=-0.35$
to $-0.65$, taking into account the velocity dispersions obtained from
spectroscopy (\S2.3) and the lens model (\S3.2).

We have computed angular position ratios of
$log(\theta_{r}/\theta_{t})=-1.04\pm 0.07$ and $-0.51\pm 0.03$ for
A\,383's inner (B0b) and outer (B1d) radial arcs respectively.  The outer
radial arc appears to be consistent with Williams et al.'s predictions,
however the inner radial arc falls well outside their prediction.
This discrepancy may be due to simplifying assumptions made by Williams
et al.\ regarding the nature and complexity of cluster substructure.
We also note that, due to the (almost) circular symmetry of A\,383, the
radial arcs are not forced to lie at the radial critical line in the
image plane.  This is another potential source of discrepancy between
the observations and the predictions.

\subsubsection{Comparison With Numerical Simulations}

Ghigna et al.\ (2000) have performed numerical simulations of galaxy
clusters at the highest resolution to date, resolving structure down to
scales of $r\sim 5$\,kpc.  They find a de-projected slope of $-1.6\pm 0.1$
on scales of $r=5$--100\,kpc in a $\sigma\sim 800$\,km\,s$^{-1}$ cluster.
This is steeper than the density profile of A\,383 on comparable scales,
$\sim -1.3$ (see \S3.2.1 and Fig.~7).  We will present a detailed
comparison of the lens model using NFW and Moore et al.\ (1998) density
profiles in a subsequent paper.

Recent progress has also been made in the inclusion of gas in 
numerical simulations (e.g.\ Pearce et al.\ 1999).  However, the 
resolution of these simulations precludes analysis of the form of 
the mass profile on the scales probed by the radial arcs in A\,383 
($r\ls 20$\,pc).  Higher resolution simulations including gas cooling 
have been performed by Lewis et al.\ (1999).  However, these authors 
concede that the mass of the central galaxies in their simulations 
is unrealistically high, which hampers a reliable comparison with our 
observations of A\,383.

The constraints on the density profile derived from the radial 
arcs in A\,383 confirm the need for a massive central galaxy to 
explain the lensing properties of this X-ray luminous cluster.  Our 
observations therefore support the case for increasing the resolution 
of N-body simulations of clusters including gas physics.

\subsection{X-ray Properties of A\,383}

The radial X-ray surface brightness profile around the cluster X-ray
centroid is shown in Fig.~8. We attempt to describe the X-ray emission
profile by fitting a standard beta model (Cavaliere \& Fusco-Femiano 1978)
\[
I(s) = I_{\rm 0} ( 1+ (\frac{r}{r_{\rm c}^{\rm 3d}})^2 
  )^{-3\beta + 0.5}\;,
\]
convolved with the {\it HRI} point-spread function (PSF), to the
observed X-ray surface brightness distribution. The free fit parameters
are the peak surface brightness ($I_{\rm 0}$), the 3-d core radius
($r_{\rm c}^{\rm 3d}$), and the slope parameter ($\beta$). We note that
the use of the beta model implies that the gas is isothermal
and in hydrostatic equilibrium. While these assumptions are likely
to be violated in most clusters, numerical simulations show that using
the beta model introduces, on average, no bias and only a
moderate scatter of typically 20\% in the cluster mass estimates
(Evrard, Metzler \& Navarro 1996).

We fit two beta models, one to the full radial range of the data including
the likely cooling flow region (model~A), and a second one to the outer
regions only (model~B). While model A yields a poor description of
the X-ray emission at large radii (see Fig.~8) it provides us with an
acceptable fit to, and an analytic description of, the emission profile
at $r\ls 100$\,kpc, which can be used to derive the gas density and,
assuming hydrostatic equilibrium, the total binding mass within the
core region (see below). Model~B, obtained by fitting only the range
from $r=12$\,kpc to 850\,kpc ($3.5'$), describes the non-cooling-flow
component (Fig.~8), for which we find $r_{\rm c}^{\rm 3d}=(23\pm 3)''$,
corresponding to $(93\pm 12)$\,kpc, and $\beta=0.65\pm 0.03$, consistent
with the canonical value of 2/3. The quoted errors include systematic
uncertainties which we explored by varying the radial range used for
the fit.

\vspace{0.2cm}
%
%
\centerline{\psfig{file=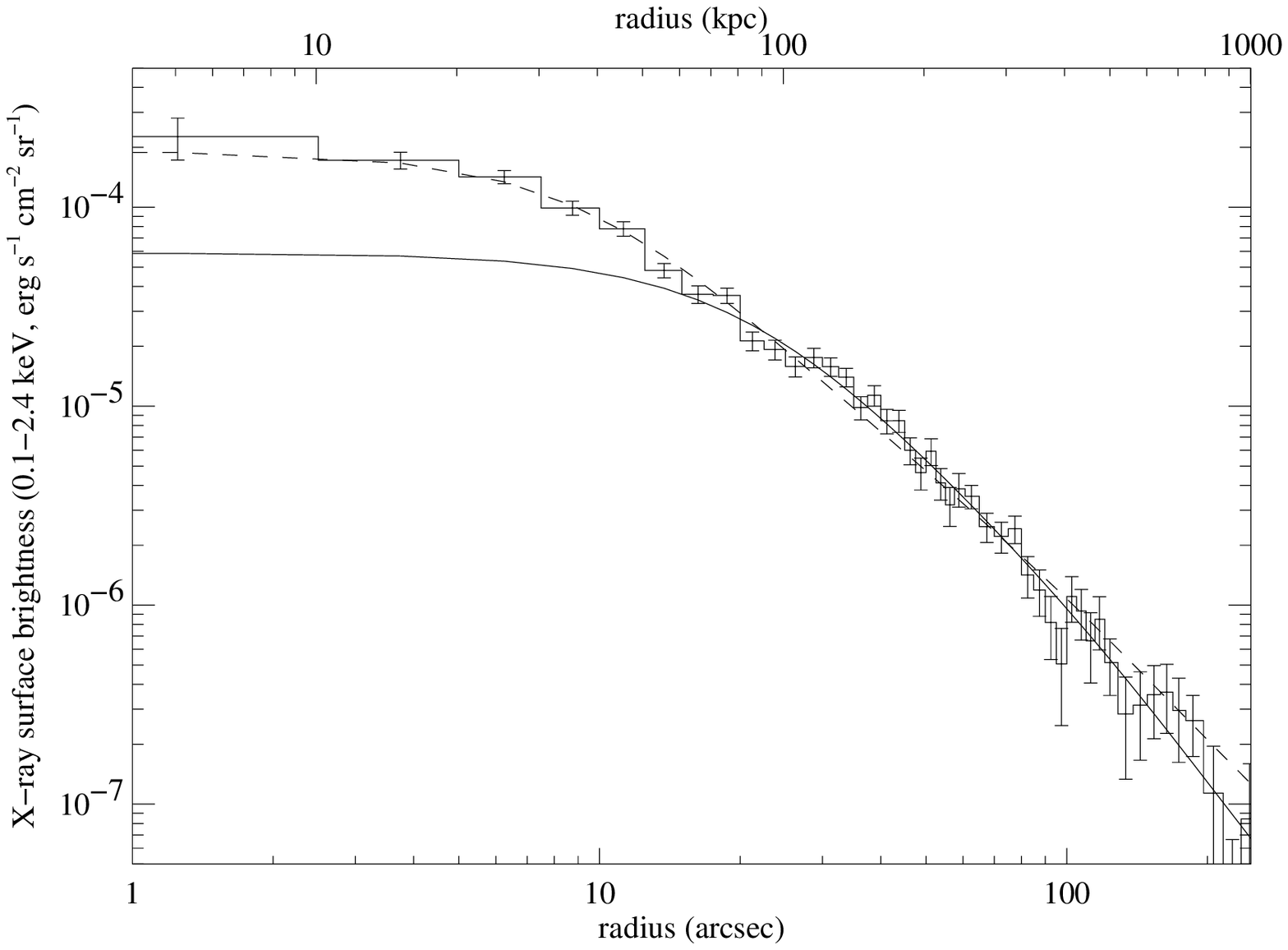,width=3.5in,angle=0}}

\noindent{\scriptsize \addtolength{\baselineskip}{-3pt} 
{\sc Fig.\,8.} -- The radial X-ray surface brightness profile of A\,383
as observed with the {\it ROSAT} {\it HRI}. Overlaid are the best-fit beta
models (convolved with the {\it HRI} PSF) and fitted to the full profile
(model~A, dashed line) or to the outer regions only (model~B, solid
line). The excess emission in the cluster core is caused by an additional
cooling flow component, which is not included in the models.

\addtolength{\baselineskip}{3pt} }
\vspace{0.2cm}

Subtracting beta model B from the observed profile we find the
cooling flow to contribute 13\% to the total cluster flux within a 
radius of 500\,kpc, corresponding to a luminosity of $1\times
10^{44}$\,erg\,s$^{-1}$ (0.1--2.4\,keV). A rough (usually low) estimate
of the mass deposition rate, $\dot{M}$, can be obtained by converting
this excess luminosity to the bolometric band (by multiplying by a factor
of 3.03 appropriate for the assumed cluster gas temperature of 7.1\,keV)
and equating it to the luminosity of the cooling flow
\[
L_{\rm cool} = \frac{5}{2}\;\frac{\dot{M}}{\mu m_{\rm p}}\; kT
\]
(Fabian 1994). We thus derive a lower limit to the mass deposition
rate of $\dot{M}\gs 200$\,$M_{\odot}$\,yr$^{-1}$, a value typical of
moderate cooling-flow clusters (e.g.\ Peres et al.\ 1998).

Under the assumption of hydrostatic equilibrium, and using a beta model
to describe the radial gas density profile, the total gravitational
mass within a projected radius $r$ can be derived from 
\[
M(<r) = \frac{3\beta\,kT\,r_{\rm c}^{\rm 3d}}{\mu\,m_{\rm p}\,G}
  \;\frac{\pi}{2}\;
  \frac{(r/r_{\rm c}^{\rm 3d})^2}{\sqrt{1+(r/r_{\rm c}^{\rm 3d})^2}}
\]
(Soucail et al.\ 2000).

Using the parameters of model~A we estimate the projected mass within
a radius of 65\,kpc to be $4.0^{+1.1}_{-1.7}\times 10^{13};M_{\odot}$,
in good agreement with our lensing mass measurement of $(3.5\pm0.1)
\times 10^{13} \,M_{\odot}$ within the same projected radius (\S3.2).
The equivalent mass estimate within a radius of 250\,kpc is more
uncertain, but is approximately $(1.2\pm 0.5) \times 10^{14} M_\odot$,
again in reasonable agreement with the lensing measurement.  We note
that the bulk of the uncertainty in the X-ray mass estimates is caused
by the adopted error of $\pm 2$\,keV in the global gas temperature,
$kT$. Forthcoming observations of our full cluster sample with {\it
Newton} will allow us to obtain much more accurate X-ray mass measurements
on scales of $\sim10$--1000\,kpc.

\section{Discussion and Conclusions}

We have discovered numerous new gravitationally lensed features,
including a giant arc and two radial arcs in A\,383, a massive, X-ray
luminous cluster of galaxies at a redshift of $z=0.188$.  This is the 
first cluster in which two radial arcs have been identified and these 
provide a detailed view of the cluster mass distribution on scales of 
$r\ls 20$\,kpc.

The morphologies and positions of the arcs have been used, in
conjunction with ground-based multi-color photometry, to constrain the
mass distribution and compute a precise mass density profile of this 
cluster within a radius of $\sim 250$\,kpc.

Within the core radius ($r=46\pm 3$\,kpc) derived from the lensing analysis,
the density profile exhibits a shallower slope ($d(\log \rho)/d(\log
r)=-1.29\pm 0.03$) than a Moore profile (Moore et al.\ 1998; Ghigna et al.\
2000) and a steeper slope than a single NFW profile.

We have also analysed the properties of the two radial arcs,
interpreting the difference in angular position of these arcs as a
signature of the lensing role of the massive central galaxy.  This
supports the proposal by Williams et al.\ (1999) that massive central
galaxy halos are required to explain the lensing properties of the 
cores of massive clusters ($\sigma\sim 1000$\,km\,s$^{-1}$).  However, 
only the outer radial arc of A\,383 falls within the range predicted 
by Williams et al.\ for the the angular position ratio of radial and 
tangential arcs.  The inner radial arc lies well outside of their 
predicted range.  This discrepancy is probably caused by simplifying 
assumptions in Williams et al.'s theoretical models.

The lensing analysis is complemented by an analysis of the cluster's
X-ray properties as obtained from archival {\it ROSAT} {\it HRI}
data.  We find strong evidence for a cooling flow in agreement with
the optical properties of the cluster, and derive a lower limit on the
mass deposition rate of 200\,M$_{\odot}$\,yr$^{-1}$, typical of moderate
cooling flow clusters.  The X-ray estimates for the total mass within the
projected radius of the giant arc are in good agreement with the lensing
measurements and are consistent with earlier claims (e.g.\ Allen 1998)
that significant discrepancies between lensing and X-ray mass estimates
are caused by substructure in unrelaxed, non-cooling-flow systems.

We note that the cluster sample used in our survey is effectively X-ray
selected, thus allowing us, for the first time, to measure in detail
the mass distribution in massive clusters of galaxies unaffected by
selection biases.  The theme of our future program will be to analyse
the lensing signal and X-ray emission (as observed with {\it Newton})
from all twelve clusters in our sample and to trace the form of the mass
profile from 50\,kpc to 5\,Mpc.  We will investigate the dispersion in
the profiles and the core properties of the clusters and determine how
these correlate with the cluster's dynamical state and the presence of
a cooling flow.  We shall also attempt to constrain the high end of the
cluster mass function and calibrate directly the cluster mass-temperature
and mass-luminosity relations at $z\sim 0.2$.

\section*{Acknowledgements}

We thank John Blakeslee, Alastair Edge, Ben Moore, Frazer Pearce and
Liliya Williams for helpful conversations and assistance.  Thanks also 
go to Laurence Jones for helping with the Keck/MOS observations.

GPS acknowledges support from PPARC.  JPK acknowledges support from
CNRS.  HE gratefully acknowledges financial support by NASA LTSA grant
NAG 5--8253.  OC acknowledges support from the European Commission
under contract no.\,ER--BFM--BI--CT97--2471.  IRS acknowledges support
from a Royal Society University Research Fellowship.  We also acknowledge 
financial support from the UK-French ALLIANCE collaboration programme 
\#00161XM.

UKIRT is operated by the Joint Astronomy Centre on behalf of the
Particle Physics and Astronomy Research Council of the United Kingdom.
The W.\,M.\ Keck Observatory is operated as a scientific partnership
among the California Institute of Technology, the University of
California and NASA.



\begin{references}
\reference{} Allen, S.W., 1998, MNRAS, 296, 392
\reference{} Allen, S.W., Fabian, A.C., 1998, MNRAS, 297, 57
\reference{} Abell, G.O., Corwin, H.G., Jr., Olowin, R.P., 1989, ApJS, 70, 1
\reference{} Bahcall, N.A., Fan, X., Cen, R., 1997, ApJ, 485, L53
\reference{} Bertin, E., Arnouts, S., 1996, A\&A, 117, 393
\reference{} Cavaliere, A., Fusco-Femiano, R., 1978, A\&A, 49, 137
\reference{} Crawford, C.S., Allen, S.W., Ebeling, H., Edge, A.C., Fabian, 
             A.C., 1999, MNRAS, 306, 857
\reference{} Czoske, O., et al., 2000, in preparation
\reference{} DeGrandi, S.\,et al., 1999, ApJ, 514, 148
\reference{} Dickey, J.M., Lockman, F.J., 1990, ARA\&A, 28, 215
\reference{} Ebeling, H., et al., 1996, MNRAS, 281, 799
\reference{} Ebeling, H., et al., 1998, MNRAS, 301, 881
\reference{} Ebeling, H., et al., 2000, MNRAS, in press
\reference{} Ebeling, H., Edge, A.C., Henry J.P., 2000, in Large Scale 
             Structure in the X-ray Universe, Proceedings of the 20-22
             September 1999 Workshop, Santorini, Greece, eds.\ Plionis, M.,
             Georgantopoulos, I., Atlantisciences, Paris, France, p.39 
\reference{} Eke, V.R., Cole, S., Frenk, C.S., 1996, MNRAS, 282, 263
\reference{} Evrard, A.E., Metzler, C.A., Navarro, J.F., 1996, ApJ, 469, 494
\reference{} Fabian, A.C., 1994, AR\&A, 32, 277
\reference{} Fahlman, G., Kaiser, N., Squires, G., Woods, D., 1994, ApJ, 
             437, 56
\reference{} Fort, B., Le F\`evre, O., Hammer F., Cailloux, M., 1992, ApJ, 399,
             L125
\reference{} Fruchter, A.S., Hook, R.N., 1997, in Applications of Digital 
             Image Processing, Proc.\,SPIE, 3164, ed.\,Tescher, A., p120
\reference{} Ghigna, S., Moore, B., Governato, F., Lake, G., Quinn, T., 
             Stadel, J., 2000, ApJ, in press
\reference{} Gioia I.M., et al.\ 1990, ApJL, 356, 35
\reference{} Holtzman, J.A., Burrows, C.J., Casertano, S., Hester, J.J., 
             Trauger, J.T., Watson, A.M., Worthey, G., 1995, PASP, 107, 1065
\reference{} Kay, S.T., Bower, R.G., 1999, MNRAS, 308, 664
\reference{} Kneib, J.-P., Mellier, Y., Fort, B., Soucail, G., Longaretti, 
             P.Y., 1994, A\&A, 286, 701
\reference{} Kneib, J.-P., Ellis, R.S., Smail, I., Couch, W.J., Sharples, 
             R.M., 1996, ApJ, 471, 643
\reference{} Lewis, G.F., Babul, A., Katz, N., Quinn, T., Hernquist, L., 
             Weinberg, D.H., 1999, astro-ph/9907097
\reference{} Mellier, Y., Fort, B., Kneib J.-P., 1993, ApJ, 407, 33
\reference{} Moore, B., Governato, F., Quinn, T., Stadel, J., Lake, G., 1998 
             ApJL, 499, 5
\reference{} Natarajan, P., Kneib, J.-P., Smail, I., Ellis, R.S., 1998,
             ApJ, 499, 600
\reference{} Navarro, J.F., Frenk, C.S., White, S.D.M., 1997, ApJ, 490, 493
\reference{} Oke, J.B., et al., 1995, PASP, 107, 375
\reference{} Pearce, F.R., Thomas, P.A., Couchman, H.M.P., Edge, A.C., 2000, 
	     MNRAS, in press
\reference{} Peres, C.B., Fabian, A.C., Edge, A.C., Allen, S.W., Johnstone, 
             R.M., White, D.A., 1998, MNRAS, 298, 416
\reference{} Smail, I., Couch, W.J., Ellis, R.S.,Sharples, R.M., 1995, ApJ, 
             440, 501
\reference{} Smail, I., et al., 1996, ApJ, 469, 508
\reference{} Smail, I., Ellis, R.S., Dressler, A., Couch, W.J., Oemler, 
             A., Butcher, H., Sharples, R.M., 1997, ApJ, 470, 70
\reference{} Soucail G., Ota, N., Boehringer, H., Czoske, O., Hattori, 
             M., Mellier, Y., 2000, A\&A, 355, 433
\reference{} Tyson, J.A., Wenk, R.A., Valdes, F., 1990, ApJ, 349, 1
\reference{} Williams, L.L.R., Navarro, J.F., Bartelmann, M., 1999, ApJ, 
             527, 535
\reference{} Viana, P.T.P., Liddle, A.R., 1996, MNRAS, 281, 323
\end{references}
\end{document}